\documentclass[nohyperref]{article}

\usepackage{microtype}
\usepackage{graphicx}
\usepackage{subfigure}
\usepackage{booktabs} 

\usepackage{hyperref}

\usepackage[accepted]{icml2022}

\usepackage{amsmath}
\usepackage{amssymb}
\usepackage{mathtools}
\usepackage{amsthm}

\usepackage[capitalize,noabbrev]{cleveref}

\theoremstyle{plain}

\theoremstyle{definition}

\theoremstyle{remark}

\usepackage[textsize=tiny]{todonotes}

\begin{document}

\twocolumn[
\icmltitle{RITA: a Study on Scaling Up Generative Protein Sequence Models}



\begin{icmlauthorlist}
\icmlauthor{Daniel Hesslow}{LightOn}
\icmlauthor{Niccol\`{o} Zanichelli}{LightOn}
\icmlauthor{Pascal Notin}{OATML}
\icmlauthor{Iacopo Poli}{LightOn}
\icmlauthor{Debora Marks}{Harvard}
\end{icmlauthorlist}

\icmlaffiliation{LightOn}{LightOn, Paris, France}
\icmlaffiliation{OATML}{Department of Computer Science, University of Oxford, Oxford, UK}
\icmlaffiliation{Harvard}{Department of Systems Biology; Harvard Medical School, Boston, MA, USA}

\icmlcorrespondingauthor{Daniel Hesslow}{\{firstname\}@lighton.ai}

\icmlkeywords{Machine Learning, ICML}

\vskip 0.3in
]



\printAffiliationsAndNotice{}  

\begin{abstract}
In this work we introduce RITA: a suite of autoregressive generative models for protein sequences, with up to 1.2 billion parameters, trained on over 280 million protein sequences belonging to the UniRef-100 database. Such generative models hold the promise of greatly accelerating protein design. We conduct the first systematic study of how capabilities evolve with model size for autoregressive transformers in the protein domain: we evaluate RITA models in next amino acid prediction, zero-shot fitness, and enzyme function prediction, showing benefits from increased scale. We release the RITA models openly, to the benefit of the research community.
\end{abstract}

\section{Introduction}
The ability to reliably design new proteins to tackle specific problems would mark the beginning of a new age, comparable to the transition between Stone and Iron age, according to \citet{designage}. While significant progress has been achieved towards this goal, beginning with directed evolution for protein engineering \cite{arnold}, much work remains to be done. 
Machine learning has been applied to a number of problems in computational biology in recent years, with promising results. A substantial fraction of recent advances in this area has been possible thanks to the application of techniques originally developed for natural language processing (NLP), in particular with the recent trend towards large language models, motivated by the discovery of scaling laws \cite{scaling_laws, chinchilla}. In the world of protein design, there have also been efforts to train large generative models.
The largest protein language model, ProGen \cite{madaniProGenLanguageModeling2020}, has been shown to be capable of generating protein sequences characterized by some desired downstream functions \cite{acrossfamilies}, but unfortunately the model remains closed source. Downstream open-source experimentation is important to discover surprising and unpredictable capabilities that are hard to discern without large-scale experimentation \cite{predictability}. This was recently exemplified when independent researchers discovered that AlphaFold 2 \cite{af2} could successfully predict multimer interactions, even though it had only been trained to predict the structure of single protein chains \cite{multimer1, multimer2}. In addition, there exists no systematic study about the evolution of capabilities with respect to model size in the protein domain: \citet{raoTransformerProteinLanguage2020} and \citet{Rives2021BiologicalSA} provided such a study for bidirectional transformers, and \citet{madaniProGenLanguageModeling2020} simply noted that their largest model was still underfitting.

Our contributions are as follows:
\begin{itemize}
    \item We introduce RITA\footnote{This project is dedicated to the loving memory of Rita Guidi (1961-2022). May ingenuity, human or otherwise, rid us one day of the disease that too soon stole you away from your loved ones.}, a family of generative protein sequence models for protein design with up to 1.2B parameters. 
    \item We study the relationship between model size and downstream task performance, taking a first step towards establishing scaling laws for protein sequence modeling.
    \item We release RITA models on Hugging Face and make them available to the scientific community at {\small \url{https://github.com/lightonai/RITA}}.
\end{itemize}

\section{Related Work}
\begin{table*}[t]
\caption{\textbf{Perplexity evaluation:} We evaluate generative protein models on the upstream modeling perplexity on four different datasets. In all cases performance is correlated with model size and RITA-XL provides the best results, highlighted in \textbf{bold}.}
\label{tab:ppl}
\vskip 0.15in
\begin{center}
\begin{small}
\begin{sc}
\begin{tabular}{lcccc c}
\toprule
& \multicolumn{4}{c}{RITA}& \multicolumn{1}{c}{Baseline}\\

Dataset & Small & Medium & Large & XLarge & ProtGPT2\\
\midrule
UniRef-100    & 10.07	& 7.47	& 6.18	& \textbf{5.48}	& 18.10\\
Metaclust     & 15.08	& 13.80	& 12.17	& \textbf{11.53}	& 21.07\\
MGnify        & 13.57	& 12.12	& 10.72	& \textbf{9.89}	& 21.10 \\
Pfam heldout  & 11.78	& 10.68	& 9.23	& \textbf{7.95}	& 15.05\\
\bottomrule
\end{tabular}
\end{sc}
\end{small}
\end{center}
\vskip -0.1in
\end{table*}

\begin{table*}[t]
\caption{\textbf{Fitness evaluation - ProteinGym substitution benchmark:} We compute the Spearman's rank correlation between the fitness value measured experimentally and the predicted fitness value across the 87 substitution DMS assays from ProteinGym, and report the average values. RITA models approach the performance of specialized models with increasing parameter count, exceeding that of ESM-1v. Baselines results are based on a single seed. Results provided in full in Table~\ref{tab:full_fitness}. Best performance is in \textbf{bold}.}
\label{tab:fitness}
\vskip 0.15in
\begin{center}
\begin{small}
\begin{sc}
\begin{tabular}{lcccc ccccc}
\toprule
& \multicolumn{4}{c}{RITA}& \multicolumn{4}{c}{Baselines} \\

& Small & Medium & Large & XLarge &  ESM-1v & MSA Transformer & Tranception & EVE \\
\midrule
AVG Fitness & 0.330 & 0.370 & 0.381 & 0.387 & 0.371 & 0.422 & \textbf{0.451} & 0.448\\
\bottomrule
\end{tabular}
\end{sc}
\end{small}
\end{center}
\vskip -0.1in
\end{table*}

\subsection{Language models for natural language processing}

Large transformer models have grown to become the de facto standard in natural language processing. As model size increased, a new paradigm of in-context learning and zero-shot classification has emerged. Instead of finetuning language models on specific tasks, models are instead trained on massive unstructured pre-training corpora where they can learn to solve a wide variety of tasks without explicit dataset curation. An important factor in the explosion of work on increasingly large language models is the discovery of scaling laws \cite{scaling_laws, chinchilla}, guiding decisions on optimal model and dataset size for a given compute budget. They allow the \textit{a priori} estimation of the expected language modeling loss, reducing the risks associated with training such large models.

While massive unstructured pre-training corpora are readily available for protein sequences, there has only been limited work in both scaling up generative protein sequence models to the sizes seen in NLP, and in studying the effect of scaling on relevant downstream tasks. For these reasons, we explore the capabilities of generative protein models as model size is increased, facilitating future work on further scaling.

\subsection{Protein Sequence Models}

Much work has gone into exploring the potential of protein sequence models. UniRep \cite{unirep} demonstrated that the internal representation learned by an LSTM-based protein sequence model was sufficient to predict protein secondary structure, stability, and downstream function. Subsequent works focusing on bidirectional models, including TAPE-BERT, ESM-1b, ProtTrans and ProteinBERT \cite{TAPE, raoTransformerProteinLanguage2020, Rives2021BiologicalSA, elnaggarProtTransCrackingLanguage2021a, brandesProteinBERTUniversalDeeplearning2021} have improved upon this by employing more capable models based on Transformers \cite{transformers}. DeepSequence \cite{deepsequence} and ESM-1v \cite{ESM1v} have shown that these representations can also be successfully leveraged for variant effect prediction, whereas RGN2 \cite{chowdhurySinglesequenceProteinStructure2021} recently demonstrated their utility for fast and accurate single-sequence tertiary structure prediction.


UniRep has been successfully employed for protein engineering \cite{unirep_design}, and the more recent transformer-based ProGen \cite{madaniProGenLanguageModeling2020} is capable of generating proteins with a number of desired characteristics by conditioning the model on a variety of sequence metadata.

Furthermore, several works have explored the use of generative protein sequence models as part of a fixed-backbone protein design pipeline, either by conditioning the sequence model on structural information through a cross-attention setup \cite{ingraham}, by coupling it with a structure predictor enhanced decoding strategy \cite{moffat_af2} or by iteratively finetuning it on sequences refined by AlphaFold 2 \cite{moffat_dark}.

\section{Methods}
\begin{table*}[t]
\caption{\textbf{Enzyme function prediction:} We predict the functional properties of proteins in SwissProt and report the top-k accuracy. Performance scales smoothly with model size and RITA-XL gives the best results, highlighted in \textbf{bold}.}
\label{tab:enz}
\vskip 0.15in
\begin{center}
\begin{small}
\begin{sc}
\begin{tabular}{lcccc c}
\toprule
& \multicolumn{4}{c}{RITA}& \multicolumn{1}{c}{Baseline}\\

top-k & Small & Medium & Large & XLarge & ProtGPT2 \\
\midrule
@1  & 87.8	& 89.5	& 90.8	& \textbf{91.6}	& 85.7\\
@3  & 90.7	& 90.2	& 93.2	& \textbf{93.7}	& 88.7\\
@10 & 92.4	& 93.9	& 94.5	& \textbf{94.7}	& 90.6 \\
\bottomrule
\end{tabular}
\end{sc}
\end{small}
\end{center}
\vskip -0.1in
\end{table*}

\subsection{Model Architecture}
A range of techniques to control neural language generation have been developed recently \cite{lillog, decodingsurvey}. However, to provide the scientific community with a model as generally applicable as possible, we chose to train our models as decoder-only transformer models without any conditioning information. We performed a small ablation study over positional embedding techniques, where we evaluated Rotary Positional Embeddings (RoPE) \cite{rotary} and AliBi \cite{alibi}, and chose to use RoPE due to the resulting lower language modeling loss, shown in Table~\ref{tab:rotary}. We trained four different models in order to study the relationship between model size and downstream capability, and use the same model hyperparameters and naming scheme as GPT-3 \citep{gpt3}.

While tokenization is widely used in natural language processing, there are important differences between natural languages and protein sequences: books consists of hundreds of thousands of characters while the average protein sequence in UniProtKB/TrEMBL is only 349 amino acids long. Protein sequences also lack a natural decomposition into something equivalent to words. Additionally, using tokenization schemes with a varied-length vocabulary may have undesirable side effects, such as producing tokenized sequences of different lengths for two proteins that would only differ by a single substitution. This would complicate the comparison of relative likelihoods, or the generation of sequences with a target output length (e.g., for fixed-backbone protein design).

\subsection{Data}

To preserve all information contained within the pre-training data we chose not to perform any clustering before training. We focus on three different pre-training corpora: UniRef-100 \cite{uniprot}, MGnify \cite{mg} and Metaclust \cite{mc}, each providing a sufficient amount of tokens for model pre-training without having to repeat the data. We then train three small models for a short amount of time to estimate transferability of each dataset to the others. The experiments showed that we would get the best results by utilizing UniRef-100 followed by Metaclust, and worst results with MGnify, as shown in Table~\ref{tab:dataset_selection}. However, we note that using a combination of several datasets may be beneficial. 

During pre-training we randomly map amino acids \texttt{B}, \texttt{Z} and \texttt{J} to \texttt{(D,N), (E,Q)} and \texttt{(I,L)} respectively and remove any sequence containing \texttt{X}. We train both on the primary sequence and its reverse.

\subsection{Training}

We utilize the Megatron-Deepspeed framework to achieve high training throughput and train the models using a combination of data and pipeline parallelism. 

All models were trained on a total of 150 billion amino acids, and the training runs were performed on the Jean Zay supercomputer of IDRIS. The models were trained for a total training time of over 25 thousand \texttt{Nvidia-V100} GPU hours. We utilize the Adam optimizer \cite{adam}, a batch size of 512, and a context size of 1024 for all experiments. 

\section{Evaluation}

\begin{table*}[t]
\caption{\textbf{Prompt tuning:} We perform prompt tuning to generate proteins from the family \texttt{PF03272}. The perplexity of the \texttt{Prompt-Tuned Model} is significantly lower than the one of the \texttt{Base Model}, indicating the the model has learned to better generate proteins from the target family.}
\label{tab:prompt}
\vskip 0.15in
\begin{center}
\begin{small}
\begin{sc}
\begin{tabular}{lcccc}
\toprule
& \multicolumn{4}{c}{RITA}\\
Perplexity & Small & Medium & Large & XLarge \\
\midrule
Base model & 15.69	& 13.60	& 10.43	& 7.35 \\
Prompt-tuned model & \textbf{10.37}	& \textbf{9.19}	& \textbf{6.99}	& \textbf{4.96} \\
\bottomrule
\end{tabular}
\end{sc}
\end{small}
\end{center}
\vskip -0.1in
\end{table*}

\subsection{Perplexity evaluation}
Autoregressive models are typically evaluated by their modeling loss. The model is trained on this task and it should broadly reflect its capabilities. We measure the perplexity on three different protein databases: UniRef-100, MGnify and Metaclust. Our models are trained on UniRef-100, a large collection of sequenced proteins, whereas Metaclust and MGnify consist of metagenomically transcribed proteins. We argue that this should provide a challenging distribution shift for the model. We additionally withheld a set of twenty protein families (the same held out by \citet{madaniProGenLanguageModeling2020}) to evaluate the generalization to unseen protein families.

For all datasets we compare our results with those of ProtGPT2 \cite{ferruz2022deep}. Since ProtGPT2, is trained on tokenized sequences, we measure the perplexity per amino acid \footnote{See appendix~\ref{sec:ppl_per_byte} for a discussion around perplexity per byte}. We present our results in Table~\ref{tab:ppl} where we can see that there is a clear improvement with increasing model size.

\subsection{Scaling Laws}
\begin{figure}
    \centering
    \includegraphics[width=0.42\textwidth]{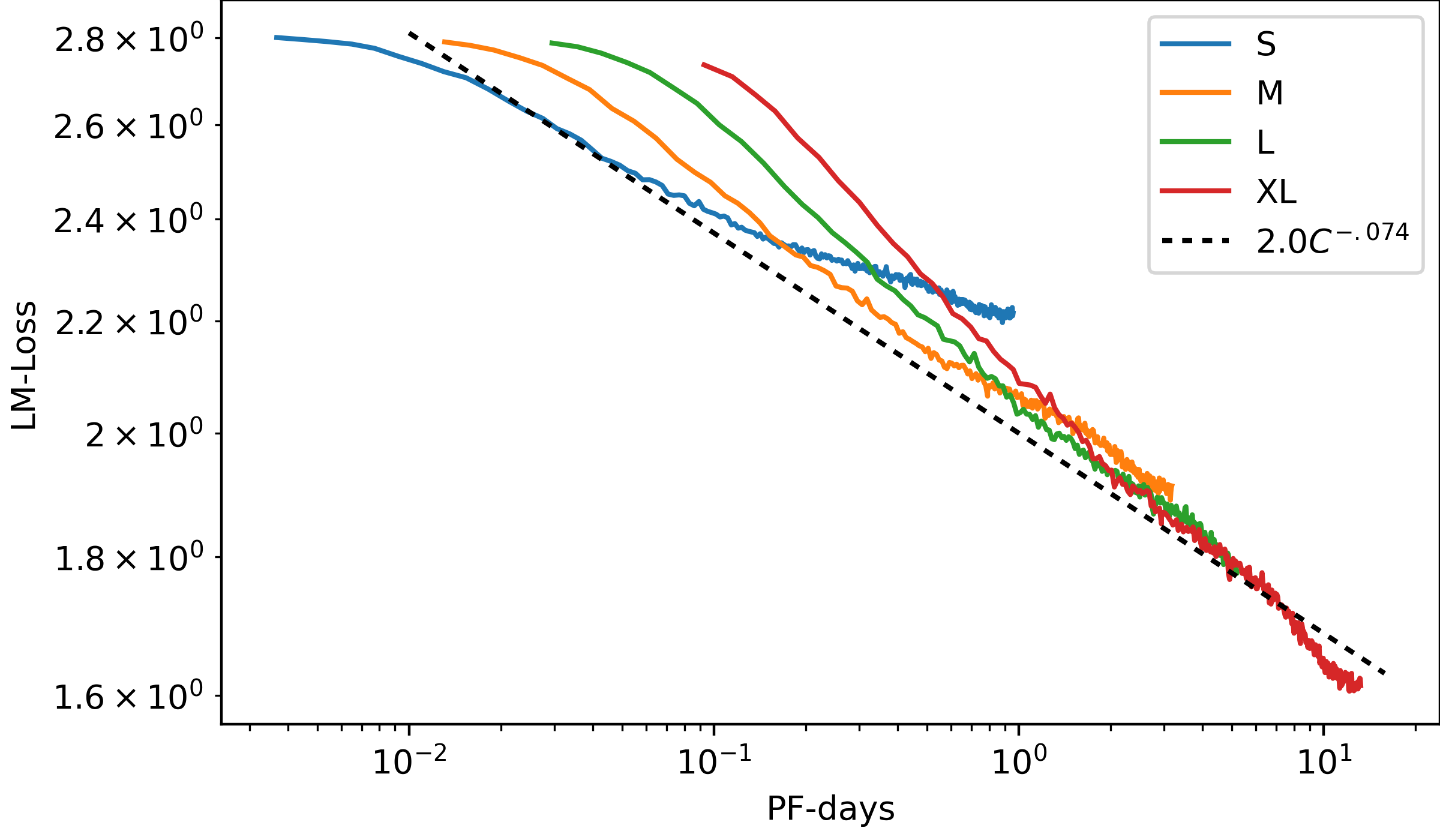}
    \caption{Protein modeling loss as a function of compute measured in PetaFLOPS-days (PF-days).}
    \label{fig:scaling_laws}
\end{figure}

By training four generative protein sequence models ranging over an order of magnitude in size, we are able to establish scaling laws similar to those established by \citet{scaling_laws} for natural language processing, see Figure~\ref{fig:scaling_laws}. We observe an exponent of $0.74$, significantly steeper than the one observed in NLP of around $0.5$. In the regime we have analyzed, up to 1.2 billion parameters, scaling up protein sequence models is thus significantly more beneficial than scaling up language models in NLP. However, in contrast to NLP, where the modeling loss typically follows the power law relationship remarkably closely, we observe some deviation. In particular, we observe a sharp decrease of loss for our largest model after a significant amount of training, and note that the loss does not go above 2.8 due to the small vocabulary size. Interestingly, even though our models are trained far beyond the point of optimality in NLP, all but our smallest model still appear to be undertrained. Our models are trained for 150 billion amino acids, whereas \citet{scaling_laws} and \citet{chinchilla} estimated optimality for our largest model at around 25 billion amino acids.

\subsection{Mutation Effects Prediction}
We assess the ability of our models to predict the effects of mutations by interpreting the likelihood that the model outputs for a given protein as its fitness value. We use the ProteinGym benchmarks \cite{Notin2022TranceptionPF} which provide experimentally-measured fitness values across 94 Deep Mutational Scanning (DMS) assays. On the substitution benchmark (Table~\ref{tab:fitness}), we compare against several baselines, including MSA Transformer \cite{raoMSATransformer2021}, ESM-1v \cite{ESM1v}, Tranception \cite{Notin2022TranceptionPF} and EVE \cite{EVE}. We observe that the performance of RITA models increases with model size, exceeding that of ESM-1v for the Large and XLarge variants. While alignment-based models (eg., EVE) and models relying on the marked-marginals heuristics for scoring (eg., ESM-1v, MSA Transformer) are unable to score indels, autoregressive transformers such as RITA models can quantify their fitness out-of-the-box, performing on par with the specialized models introduced in \citet{shinProteinDesignVariant2021} (Table~\ref{tab:full_fitness_indels}).

\subsection{Enzyme Function Prediction}
To evaluate the capability of the models to predict enzyme function, we utilize the sequence representation obtained at the final token. Following ProteInfer \cite{proteinfer}, we extract enzyme commission metadata from SwissProt \cite{swissprot} and focus on the tags belonging to the finest classification level (randomly choosing one for those with multiple tags), obtaining a classification problem with 4793 classes. The processed dataset is available at
{\small{\url{https://huggingface.co/datasets/lightonai/SwissProt-EC-leaf}}}.

We train a linear classifier for one epoch on top of the extracted representations, and present the results in Table~\ref{tab:enz}. Similar to previous tasks, performance increases with scale.

\subsection{Prompt Tuning}
Language models can solve a wide variety of tasks by prefixing the generation with a manually created prompt. This practice has received the name of \textit{prompt engineering}, in reference to the laborious process of finding a prompt that yields satisfactory generation. Inspired by this, \citet{prompt_tuning} developed \textit{prompt tuning}, a method to automatically learn soft prompts in the embedding space. Prompt tuning has emerged as an important way to perform \textit{parameter efficient fine-tuning}, learning only a fraction of the number of parameters typically needed for fine-tuning. 

We investigate if it is possible to add controllable generation to pre-trained protein sequence models by leveraging \textit{prompt tuning}. We arbitrarily chose one of the protein families that were held out during training, \texttt{PF03272}, and learned a prompt specializing in generating proteins from this family. In Table~\ref{tab:prompt} we see a significant reduction in perplexity with prompt tuning, showing that the model is indeed able to learn to generate proteins from this protein family. 

\section{Conclusion and Future Work}

In this work we have presented RITA, a family of generative protein sequence models, aiming to accelerate future work on protein design. We have systematically evaluated how model capabilities increase with size and taken a first step towards establishing scaling laws for protein sequence modeling. We believe that the release of our models will represent a building block for future endeavours into protein design. We also look forward to future work studying RITA-designed proteins \textit{in-vitro}, further scaling up protein sequence models or augmenting them with target structure embeddings for fixed-backbone protein design. 

\newpage

\bibliography{citations}

\begin{thebibliography}{41}
\providecommand{\natexlab}[1]{#1}
\providecommand{\url}[1]{\texttt{#1}}
\expandafter\ifx\csname urlstyle\endcsname\relax
  \providecommand{\doi}[1]{doi: #1}\else
  \providecommand{\doi}{doi: \begingroup \urlstyle{rm}\Url}\fi

\bibitem[Alley et~al.(2019)Alley, Khimulya, Biswas, AlQuraishi, and
  Church]{unirep}
Alley, E.~C., Khimulya, G., Biswas, S., AlQuraishi, M., and Church, G.~M.
\newblock Unified rational protein engineering with sequence-only deep
  representation learning.
\newblock \emph{bioRxiv}, 2019.
\newblock \doi{10.1101/589333}.
\newblock URL \url{https://www.biorxiv.org/content/early/2019/03/26/589333}.

\bibitem[Arnold(1998)]{arnold}
Arnold, F.~H.
\newblock Design by directed evolution.
\newblock \emph{Accounts of Chemical Research}, 31:\penalty0 125--131, 1998.

\bibitem[Baek(2021)]{multimer2}
Baek, M.
\newblock Twitter post: Adding a big enough number for residue index feature is
  enough to model hetero-complex using alphafold (green\&cyan\: crystal
  structure \/ magenta: predicted model w/ residue\_index modification), July
  2021.
\newblock URL \url{https://twitter.com/minkbaek/status/1417538291709071362}.

\bibitem[Bairoch \& Apweiler(2000)Bairoch and Apweiler]{swissprot}
Bairoch, A. and Apweiler, R.
\newblock The swiss-prot protein sequence database and its supplement trembl in
  2000.
\newblock \emph{Nucleic acids research}, 28\penalty0 (1):\penalty0 45--48,
  2000.

\bibitem[Biswas et~al.(2020)Biswas, Khimulya, Alley, Esvelt, and
  Church]{unirep_design}
Biswas, S., Khimulya, G., Alley, E.~C., Esvelt, K.~M., and Church, G.~M.
\newblock Low-n protein engineering with data-efficient deep learning.
\newblock \emph{bioRxiv}, 2020.
\newblock \doi{10.1101/2020.01.23.917682}.
\newblock URL
  \url{https://www.biorxiv.org/content/early/2020/01/24/2020.01.23.917682}.

\bibitem[Brandes et~al.(2021)Brandes, Ofer, Peleg, Rappoport, and
  Linial]{brandesProteinBERTUniversalDeeplearning2021}
Brandes, N., Ofer, D., Peleg, Y., Rappoport, N., and Linial, M.
\newblock {{ProteinBERT}}: {{A}} universal deep-learning model of protein
  sequence and function, May 2021.

\bibitem[Brown et~al.(2020)Brown, Mann, Ryder, Subbiah, Kaplan, Dhariwal,
  Neelakantan, Shyam, Sastry, Askell, et~al.]{gpt3}
Brown, T., Mann, B., Ryder, N., Subbiah, M., Kaplan, J.~D., Dhariwal, P.,
  Neelakantan, A., Shyam, P., Sastry, G., Askell, A., et~al.
\newblock Language models are few-shot learners.
\newblock \emph{Advances in neural information processing systems},
  33:\penalty0 1877--1901, 2020.

\bibitem[Chowdhury et~al.(2021)Chowdhury, Bouatta, Biswas, Rochereau, Church,
  Sorger, and AlQuraishi]{chowdhurySinglesequenceProteinStructure2021}
Chowdhury, R., Bouatta, N., Biswas, S., Rochereau, C., Church, G.~M., Sorger,
  P.~K., and AlQuraishi, M.
\newblock Single-sequence protein structure prediction using language models
  from deep learning, August 2021.

\bibitem[Elnaggar et~al.(2021)Elnaggar, Heinzinger, Dallago, Rehawi, Wang,
  Jones, Gibbs, Feher, Angerer, Steinegger, Bhowmik, and
  Rost]{elnaggarProtTransCrackingLanguage2021a}
Elnaggar, A., Heinzinger, M., Dallago, C., Rehawi, G., Wang, Y., Jones, L.,
  Gibbs, T., Feher, T., Angerer, C., Steinegger, M., Bhowmik, D., and Rost, B.
\newblock {{ProtTrans}}: {{Towards Cracking}} the {{Language}} of {{Lifes Code
  Through Self-Supervised Deep Learning}} and {{High Performance Computing}}.
\newblock \emph{IEEE Transactions on Pattern Analysis and Machine
  Intelligence}, pp.\  1--1, 2021.
\newblock ISSN 0162-8828, 2160-9292, 1939-3539.
\newblock \doi{10.1109/TPAMI.2021.3095381}.

\bibitem[Ferruz et~al.(2022)Ferruz, Schmidt, and H{\"o}cker]{ferruz2022deep}
Ferruz, N., Schmidt, S., and H{\"o}cker, B.
\newblock A deep unsupervised language model for protein design.
\newblock \emph{bioRxiv}, 2022.

\bibitem[Frazer et~al.(2021)Frazer, Notin, Dias, Gomez, Min, Brock, Gal, and
  Marks]{EVE}
Frazer, J., Notin, P., Dias, M., Gomez, A.~N., Min, J.~K., Brock, K.~P., Gal,
  Y., and Marks, D.~S.
\newblock Disease variant prediction with deep generative models of
  evolutionary data.
\newblock \emph{Nature}, 2021.

\bibitem[Ganguli et~al.(2022)Ganguli, Hernandez, Lovitt, DasSarma, Henighan,
  Jones, Joseph, Kernion, Mann, Askell, Bai, Chen, Conerly, Drain, Elhage,
  Showk, Fort, Hatfield-Dodds, Johnston, Kravec, Nanda, Ndousse, Olsson,
  Amodei, Amodei, Brown, Kaplan, McCandlish, Olah, and Clark]{predictability}
Ganguli, D., Hernandez, D., Lovitt, L., DasSarma, N., Henighan, T.~J., Jones,
  A., Joseph, N., Kernion, J., Mann, B., Askell, A., Bai, Y., Chen, A.,
  Conerly, T., Drain, D., Elhage, N., Showk, S.~E., Fort, S., Hatfield-Dodds,
  Z., Johnston, S., Kravec, S., Nanda, N., Ndousse, K., Olsson, C., Amodei, D.,
  Amodei, D., Brown, T.~B., Kaplan, J., McCandlish, S., Olah, C., and Clark, J.
\newblock Predictability and surprise in large generative models.
\newblock \emph{ArXiv}, abs/2202.07785, 2022.

\bibitem[Hoffmann et~al.(2022)Hoffmann, Borgeaud, Mensch, Buchatskaya, Cai,
  Rutherford, de~Las~Casas, Hendricks, Welbl, Clark, Hennigan, Noland,
  Millican, van~den Driessche, Damoc, Guy, Osindero, Simonyan, Elsen, Rae,
  Vinyals, and Sifre]{chinchilla}
Hoffmann, J., Borgeaud, S., Mensch, A., Buchatskaya, E., Cai, T., Rutherford,
  E., de~Las~Casas, D., Hendricks, L.~A., Welbl, J., Clark, A., Hennigan, T.,
  Noland, E., Millican, K., van~den Driessche, G., Damoc, B., Guy, A.,
  Osindero, S., Simonyan, K., Elsen, E., Rae, J.~W., Vinyals, O., and Sifre, L.
\newblock Training compute-optimal large language models.
\newblock \emph{ArXiv}, abs/2203.15556, 2022.

\bibitem[Huang et~al.(2016)Huang, Boyken, and Baker]{designage}
Huang, P.-S., Boyken, S.~E., and Baker, D.
\newblock The coming of age of de novo protein design.
\newblock \emph{Nature}, 537:\penalty0 320--327, 2016.

\bibitem[Ingraham et~al.(2019)Ingraham, Garg, Barzilay, and Jaakkola]{ingraham}
Ingraham, J., Garg, V., Barzilay, R., and Jaakkola, T.
\newblock Generative models for graph-based protein design.
\newblock In Wallach, H., Larochelle, H., Beygelzimer, A., d\textquotesingle
  Alch\'{e}-Buc, F., Fox, E., and Garnett, R. (eds.), \emph{Advances in Neural
  Information Processing Systems}, volume~32. Curran Associates, Inc., 2019.
\newblock URL
  \url{https://proceedings.neurips.cc/paper/2019/file/f3a4ff4839c56a5f460c88cce3666a2b-Paper.pdf}.

\bibitem[Jumper et~al.(2021)Jumper, Evans, Pritzel, Green, Figurnov,
  Ronneberger, Tunyasuvunakool, Bates, Z{\'i}dek, Potapenko, Bridgland, Meyer,
  Kohl, Ballard, Cowie, Romera-Paredes, Nikolov, Jain, Adler, Back, Petersen,
  Reiman, Clancy, Zielinski, Steinegger, Pacholska, Berghammer, Bodenstein,
  Silver, Vinyals, Senior, Kavukcuoglu, Kohli, and Hassabis]{af2}
Jumper, J.~M., Evans, R., Pritzel, A., Green, T., Figurnov, M., Ronneberger,
  O., Tunyasuvunakool, K., Bates, R., Z{\'i}dek, A., Potapenko, A., Bridgland,
  A., Meyer, C., Kohl, S. A.~A., Ballard, A., Cowie, A., Romera-Paredes, B.,
  Nikolov, S., Jain, R., Adler, J., Back, T., Petersen, S., Reiman, D.~A.,
  Clancy, E., Zielinski, M., Steinegger, M., Pacholska, M., Berghammer, T.,
  Bodenstein, S., Silver, D., Vinyals, O., Senior, A.~W., Kavukcuoglu, K.,
  Kohli, P., and Hassabis, D.
\newblock Highly accurate protein structure prediction with alphafold.
\newblock \emph{Nature}, 596:\penalty0 583 -- 589, 2021.

\bibitem[Kaplan et~al.(2020)Kaplan, McCandlish, Henighan, Brown, Chess, Child,
  Gray, Radford, Wu, and Amodei]{scaling_laws}
Kaplan, J., McCandlish, S., Henighan, T., Brown, T.~B., Chess, B., Child, R.,
  Gray, S., Radford, A., Wu, J., and Amodei, D.
\newblock Scaling laws for neural language models.
\newblock \emph{arXiv preprint arXiv:2001.08361}, 2020.

\bibitem[Kingma \& Ba(2015)Kingma and Ba]{adam}
Kingma, D.~P. and Ba, J.
\newblock Adam: A method for stochastic optimization.
\newblock \emph{CoRR}, abs/1412.6980, 2015.

\bibitem[Lester et~al.(2021)Lester, Al-Rfou, and Constant]{prompt_tuning}
Lester, B., Al-Rfou, R., and Constant, N.
\newblock The power of scale for parameter-efficient prompt tuning.
\newblock \emph{arXiv preprint arXiv:2104.08691}, 2021.

\bibitem[Madani et~al.(2020)Madani, McCann, Naik, Keskar, Anand, Eguchi, Huang,
  and Socher]{madaniProGenLanguageModeling2020}
Madani, A., McCann, B., Naik, N., Keskar, N.~S., Anand, N., Eguchi, R.~R.,
  Huang, P.-S., and Socher, R.
\newblock {{ProGen}}: {{Language Modeling}} for {{Protein Generation}}.
\newblock \emph{arXiv:2004.03497 [cs, q-bio, stat]}, March 2020.

\bibitem[Madani et~al.(2021)Madani, Krause, Greene, Subramanian, Mohr, Holton,
  Olmos, Xiong, Sun, Socher, Fraser, and Naik]{acrossfamilies}
Madani, A., Krause, B., Greene, E.~R., Subramanian, S., Mohr, B.~P., Holton,
  J.~M., Olmos, J.~L., Xiong, C., Sun, Z.~Z., Socher, R., Fraser, J.~S., and
  Naik, N.
\newblock Deep neural language modeling enables functional protein generation
  across families.
\newblock \emph{bioRxiv}, 2021.

\bibitem[Meier et~al.(2021)Meier, Rao, Verkuil, Liu, Sercu, and Rives]{ESM1v}
Meier, J., Rao, R., Verkuil, R., Liu, J., Sercu, T., and Rives, A.
\newblock Language models enable zero-shot prediction of the effects of
  mutations on protein function.
\newblock \emph{bioRxiv}, 2021.
\newblock \doi{10.1101/2021.07.09.450648}.
\newblock URL
  \url{https://www.biorxiv.org/content/early/2021/11/17/2021.07.09.450648}.

\bibitem[Mitchell et~al.(2020)Mitchell, Almeida, Beracochea, Boland, Burgin,
  Cochrane, Crusoe, Kale, Potter, Richardson, Sakharova, Scheremetjew,
  Korobeynikov, Shlemov, Kunyavskaya, Lapidus, and Finn]{mg}
Mitchell, A.~L., Almeida, A., Beracochea, M., Boland, M.~A., Burgin, J.,
  Cochrane, G., Crusoe, M.~R., Kale, V., Potter, S.~C., Richardson, L.~J.,
  Sakharova, E.~A., Scheremetjew, M., Korobeynikov, A.~I., Shlemov, A.,
  Kunyavskaya, O., Lapidus, A.~L., and Finn, R.~D.
\newblock Mgnify: the microbiome analysis resource in 2020.
\newblock \emph{Nucleic Acids Research}, 48:\penalty0 D570 -- D578, 2020.

\bibitem[Moffat et~al.(2021)Moffat, Greener, and Jones]{moffat_af2}
Moffat, L., Greener, J.~G., and Jones, D.~T.
\newblock Using alphafold for rapid and accurate fixed backbone protein design.
\newblock \emph{bioRxiv}, 2021.
\newblock \doi{10.1101/2021.08.24.457549}.
\newblock URL
  \url{https://www.biorxiv.org/content/early/2021/08/26/2021.08.24.457549}.

\bibitem[Moffat et~al.(2022)Moffat, Kandathil, and Jones]{moffat_dark}
Moffat, L., Kandathil, S.~M., and Jones, D.~T.
\newblock Design in the dark: Learning deep generative models for de novo
  protein design.
\newblock \emph{bioRxiv}, 2022.
\newblock \doi{10.1101/2022.01.27.478087}.
\newblock URL
  \url{https://www.biorxiv.org/content/early/2022/01/28/2022.01.27.478087}.

\bibitem[Notin et~al.(2022)Notin, Dias, Frazer, Marchena-Hurtado, Gomez, Marks,
  and Gal]{Notin2022TranceptionPF}
Notin, P., Dias, M., Frazer, J., Marchena-Hurtado, J., Gomez, A.~N., Marks,
  D.~S., and Gal, Y.
\newblock Tranception: protein fitness prediction with autoregressive
  transformers and inference-time retrieval.
\newblock \emph{ArXiv}, abs/2205.13760, 2022.

\bibitem[Press et~al.(2021)Press, Smith, and Lewis]{alibi}
Press, O., Smith, N.~A., and Lewis, M.
\newblock Train short, test long: Attention with linear biases enables input
  length extrapolation.
\newblock \emph{ArXiv}, abs/2108.12409, 2021.

\bibitem[Rao et~al.(2019)Rao, Bhattacharya, Thomas, Duan, Chen, Canny, Abbeel,
  and Song]{TAPE}
Rao, R., Bhattacharya, N., Thomas, N., Duan, Y., Chen, X., Canny, J., Abbeel,
  P., and Song, Y.~S.
\newblock Evaluating protein transfer learning with tape, 2019.
\newblock URL \url{https://arxiv.org/abs/1906.08230}.

\bibitem[Rao et~al.(2020)Rao, Meier, Sercu, Ovchinnikov, and
  Rives]{raoTransformerProteinLanguage2020}
Rao, R., Meier, J., Sercu, T., Ovchinnikov, S., and Rives, A.
\newblock Transformer protein language models are unsupervised structure
  learners, December 2020.

\bibitem[Rao et~al.(2021)Rao, Liu, Verkuil, Meier, Canny, Abbeel, Sercu, and
  Rives]{raoMSATransformer2021}
Rao, R., Liu, J., Verkuil, R., Meier, J., Canny, J.~F., Abbeel, P., Sercu, T.,
  and Rives, A.
\newblock {{MSA Transformer}}, August 2021.

\bibitem[Riesselman et~al.(2017)Riesselman, Ingraham, and Marks]{deepsequence}
Riesselman, A.~J., Ingraham, J.~B., and Marks, D.~S.
\newblock Deep generative models of genetic variation capture mutation effects.
\newblock \emph{arXiv preprint arXiv:1712.06527}, 2017.

\bibitem[Rives et~al.(2021)Rives, Meier, Sercu, Goyal, Lin, Liu, Guo, Ott,
  Zitnick, Ma, and Fergus]{Rives2021BiologicalSA}
Rives, A., Meier, J., Sercu, T., Goyal, S., Lin, Z., Liu, J., Guo, D., Ott, M.,
  Zitnick, C.~L., Ma, J., and Fergus, R.
\newblock Biological structure and function emerge from scaling unsupervised
  learning to 250 million protein sequences.
\newblock \emph{Proceedings of the National Academy of Sciences of the United
  States of America}, 118, 2021.

\bibitem[Sanderson et~al.(2021)Sanderson, Bileschi, Belanger, and
  Colwell]{proteinfer}
Sanderson, T., Bileschi, M.~L., Belanger, D., and Colwell, L.
\newblock Proteinfer: deep networks for protein functional inference.
\newblock \emph{bioRxiv}, 2021.

\bibitem[Shin et~al.(2021)Shin, Riesselman, Kollasch, McMahon, Simon, Sander,
  Manglik, Kruse, and Marks]{shinProteinDesignVariant2021}
Shin, J.-E., Riesselman, A.~J., Kollasch, A.~W., McMahon, C., Simon, E.,
  Sander, C., Manglik, A., Kruse, A.~C., and Marks, D.~S.
\newblock Protein design and variant prediction using autoregressive generative
  models.
\newblock \emph{Nature Communications}, 12\penalty0 (1):\penalty0 2403, April
  2021.
\newblock ISSN 2041-1723.
\newblock \doi{10.1038/s41467-021-22732-w}.

\bibitem[Steinegger \& S{\"o}ding(2018)Steinegger and S{\"o}ding]{mc}
Steinegger, M. and S{\"o}ding, J.
\newblock Clustering huge protein sequence sets in linear time.
\newblock \emph{Nature Communications}, 9, 2018.

\bibitem[Su et~al.(2021)Su, Lu, Pan, Wen, and Liu]{rotary}
Su, J., Lu, Y., Pan, S., Wen, B., and Liu, Y.
\newblock Roformer: Enhanced transformer with rotary position embedding.
\newblock \emph{ArXiv}, abs/2104.09864, 2021.

\bibitem[{The UniProt Consortium}(2020)]{uniprot}
{The UniProt Consortium}.
\newblock {UniProt: the universal protein knowledgebase in 2021}.
\newblock \emph{Nucleic Acids Research}, 49\penalty0 (D1):\penalty0 D480--D489,
  11 2020.
\newblock ISSN 0305-1048.
\newblock \doi{10.1093/nar/gkaa1100}.
\newblock URL \url{https://doi.org/10.1093/nar/gkaa1100}.

\bibitem[Vaswani et~al.(2017)Vaswani, Shazeer, Parmar, Uszkoreit, Jones, Gomez,
  Kaiser, and Polosukhin]{transformers}
Vaswani, A., Shazeer, N., Parmar, N., Uszkoreit, J., Jones, L., Gomez, A.~N.,
  Kaiser, L.~u., and Polosukhin, I.
\newblock Attention is all you need.
\newblock In Guyon, I., Luxburg, U.~V., Bengio, S., Wallach, H., Fergus, R.,
  Vishwanathan, S., and Garnett, R. (eds.), \emph{Advances in Neural
  Information Processing Systems}, volume~30. Curran Associates, Inc., 2017.
\newblock URL
  \url{https://proceedings.neurips.cc/paper/2017/file/3f5ee243547dee91fbd053c1c4a845aa-Paper.pdf}.

\bibitem[Weng(2021)]{lillog}
Weng, L.
\newblock Controllable neural text generation.
\newblock \emph{lilianweng.github.io}, 2021.
\newblock URL
  \url{https://lilianweng.github.io/posts/2021-01-02-controllable-text-generation/}.

\bibitem[Yoshitaka(2021)]{multimer1}
Yoshitaka, M.
\newblock Twitter post: Alphafold2 can also predict heterocomplexes. all you
  have to do is input the two sequences you want to predict and connect them
  with a long linker., July 2021.
\newblock URL \url{hhttps://twitter.com/Ag_smith/status/1417063635000598528}.

\bibitem[Zarrie{\ss} et~al.(2021)Zarrie{\ss}, Voigt, and
  Sch{\"u}z]{decodingsurvey}
Zarrie{\ss}, S., Voigt, H., and Sch{\"u}z, S.
\newblock Decoding methods in neural language generation: A survey.
\newblock \emph{Inf.}, 12:\penalty0 355, 2021.

\end{thebibliography}
\bibliographystyle{icml2022}

\newpage
\appendix
\onecolumn
\section{Full results from our fitness evaluation.}

\begin{table}[H]
\caption{\textbf{Fitness evaluation - ProteinGym substitution benchmark:} Spearman's rank correlation between experimentally measured fitness values for different proteins and the value predicted by the models. Baselines results are based on a single seed. Tranception NR and Tranception R are variants without and with retrieval respectively. We follow the same approach as in \citet{Notin2022TranceptionPF} and aggregate results at the Uniprot ID level to avoid biasing results towards proteins for which several assays are available. To compare with models relying on evolutionary data in the form of multiple-sequence alignments (eg., EVE), we only evaluate on the subset of mutations where coverage is deemed high enough by these models to make a prediction.}
\begin{center}
\begin{tiny}
\begin{tabular}{lccccccccc}
\toprule
Uniprot\_ID & \multicolumn{4}{c}{RITA}& \multicolumn{3}{c}{Baselines} \\
                                                 & \multicolumn{1}{c}{S} & \multicolumn{1}{c}{M} & \multicolumn{1}{c}{L} & \multicolumn{1}{c}{XL} & \multicolumn{1}{l}{ESM-1v} & \multicolumn{1}{l}{MSA Transformer} &
                                                 \multicolumn{1}{l}{Tranception NR} &
                                                 \multicolumn{1}{l}{Tranception R} &
                                                 \multicolumn{1}{l}{EVE} \\
\midrule
\texttt{A0A140D2T1\_ZIKV} & 0.350 & 0.308 & 0.317 & 0.304 & -0.064 & 0.465 & 0.268 & 0.346 & 0.366\\
\texttt{A0A192B1T2\_9HIV1} & 0.492 & 0.504 & 0.504 & 0.501 & 0.488 & 0.510 & 0.510 & 0.509 & 0.510\\
\texttt{A0A1I9GEU1\_NEIME} & -0.022 & 0.028 & 0.061 & 0.074 & 0.046 & 0.077 & 0.088 & 0.044 & -0.004\\
\texttt{A0A2Z5U3Z0\_9INFA} & 0.456 & 0.518 & 0.502 & 0.525 & 0.485 & 0.326 & 0.528 & 0.545 & 0.529\\
\texttt{A4D664\_9INFA} & 0.329 & 0.386 & 0.404 & 0.398 & 0.026 & 0.333 & 0.404 & 0.393 & 0.409\\
\texttt{A4GRB6\_PSEAI} & 0.411 & 0.537 & 0.562 & 0.619 & 0.647 & 0.707 & 0.598 & 0.663 & 0.672\\
\texttt{A4\_HUMAN} & 0.322 & 0.276 & 0.312 & 0.300 & 0.309 & 0.394 & 0.364 & 0.452 & 0.301\\
\texttt{AACC1\_PSEAI} & 0.271 & 0.292 & 0.349 & 0.402 & 0.488 & 0.505 & 0.407 & 0.448 & 0.499\\
\texttt{ADRB2\_HUMAN} & 0.514 & 0.511 & 0.513 & 0.484 & 0.523 & 0.436 & 0.501 & 0.542 & 0.534\\
\texttt{AMIE\_PSEAE} & 0.488 & 0.517 & 0.532 & 0.563 & 0.606 & 0.611 & 0.501 & 0.585 & 0.558\\
\texttt{B3VI55\_LIPST} & 0.284 & 0.389 & 0.431 & 0.454 & 0.483 & 0.535 & 0.491 & 0.468 & 0.436\\
\texttt{BLAT\_ECOLX} & 0.564 & 0.556 & 0.546 & 0.524 & 0.646 & 0.681 & 0.489 & 0.627 & 0.682\\
\texttt{BRCA1\_HUMAN} & 0.389 & 0.397 & 0.495 & 0.499 & 0.442 & 0.402 & 0.538 & 0.574 & 0.322\\
\texttt{C6KNH7\_9INFA} & 0.394 & 0.369 & 0.371 & 0.373 & 0.420 & 0.408 & 0.401 & 0.445 & 0.435\\
\texttt{CALM1\_HUMAN} & 0.186 & 0.245 & 0.258 & 0.275 & 0.246 & 0.254 & 0.306 & 0.283 & 0.244\\
\texttt{CAPSD\_AAV2S} & 0.191 & 0.250 & 0.269 & 0.279 & 0.196 & 0.350 & 0.492 & 0.473 & 0.346\\
\texttt{CCDB\_ECOLI} & 0.142 & 0.076 & 0.029 & 0.210 & 0.426 & 0.489 & 0.309 & 0.456 & 0.490\\
\texttt{CP2C9\_HUMAN} & 0.596 & 0.593 & 0.603 & 0.581 & 0.623 & 0.596 & 0.595 & 0.652 & 0.625\\
\texttt{DLG4\_HUMAN} & 0.573 & 0.578 & 0.563 & 0.535 & 0.544 & 0.595 & 0.576 & 0.662 & 0.594\\
\texttt{DLG4\_RAT} & 0.381 & 0.390 & 0.377 & 0.371 & 0.565 & 0.507 & 0.304 & 0.446 & 0.523\\
\texttt{DYR\_ECOLI} & 0.198 & 0.361 & 0.267 & 0.313 & 0.420 & 0.488 & 0.348 & 0.424 & 0.468\\
\texttt{ENV\_HV1B9} & 0.380 & 0.358 & 0.408 & 0.419 & 0.415 & 0.380 & 0.404 & 0.407 & 0.388\\
\texttt{ENV\_HV1BR} & 0.352 & 0.36 & 0.372 & 0.364 & 0.322 & 0.345 & 0.358 & 0.363 & 0.341\\
\texttt{ESTA\_BACSU} & 0.122 & 0.199 & 0.283 & 0.301 & 0.304 & 0.428 & 0.263 & 0.325 & 0.375\\
\texttt{F7YBW8\_MESOW} & -0.076 & -0.123 & -0.105 & -0.006 & 0.382 & 0.375 & 0.434 & 0.425 & 0.411\\
\texttt{GAL4\_YEAST} & 0.287 & 0.345 & 0.356 & 0.353 & 0.441 & 0.583 & 0.326 & 0.526 & 0.511\\
\texttt{GCN4\_YEAST} & 0.385 & 0.417 & 0.411 & 0.407 & 0.288 & 0.288 & 0.384 & 0.356 & 0.252\\
\texttt{GFP\_AEQVI} & 0.080 & 0.108 & 0.182 & 0.096 & 0.099 & 0.652 & 0.631 & 0.677 & 0.679\\
\texttt{GRB2\_HUMAN} & 0.522 & 0.471 & 0.484 & 0.382 & 0.484 & 0.468 & 0.429 & 0.489 & 0.566\\
\texttt{HIS7\_YEAST} & 0.325 & 0.402 & 0.433 & 0.478 & 0.411 & 0.508 & 0.585 & 0.616 & 0.476\\
\texttt{HSP82\_YEAST} & 0.432 & 0.439 & 0.427 & 0.457 & 0.500 & 0.445 & 0.436 & 0.461 & 0.469\\
\texttt{I6TAH8\_I68A0} & 0.308 & 0.328 & 0.374 & 0.377 & 0.018 & 0.303 & 0.337 & 0.348 & 0.364\\
\texttt{IF1\_ECOLI} & 0.364 & 0.459 & 0.381 & 0.417 & 0.538 & 0.227 & 0.548 & 0.509 & 0.525\\
\texttt{KCNH2\_HUMAN} & 0.452 & 0.489 & 0.473 & 0.434 & 0.233 & 0.368 & 0.484 & 0.513 & 0.229\\
\texttt{KKA2\_KLEPN} & 0.296 & 0.425 & 0.538 & 0.556 & 0.614 & 0.576 & 0.584 & 0.584 & 0.597\\
\texttt{MK01\_HUMAN} & 0.220 & 0.129 & 0.099 & 0.053 & 0.183 & 0.153 & 0.034 & 0.139 & 0.251\\
\texttt{MSH2\_HUMAN} & 0.303 & 0.325 & 0.278 & 0.263 & 0.398 & 0.410 & 0.292 & 0.360 & 0.405\\
\texttt{MTH3\_HAEAE} & 0.358 & 0.491 & 0.625 & 0.677 & 0.701 & 0.687 & 0.673 & 0.655 & 0.710\\
\texttt{NCAP\_I34A1} & 0.352 & 0.382 & 0.408 & 0.413 & 0.019 & 0.338 & 0.415 & 0.424 & 0.363\\
\texttt{NRAM\_I33A0} & 0.583 & 0.633 & 0.584 & 0.571 & 0.162 & 0.519 & 0.551 & 0.621 & 0.584\\
\texttt{NUD15\_HUMAN} & 0.316 & 0.451 & 0.513 & 0.498 & 0.615 & 0.630 & 0.547 & 0.591 & 0.608\\
\texttt{P53\_HUMAN} & 0.364 & 0.484 & 0.478 & 0.448 & 0.487 & 0.396 & 0.388 & 0.461 & 0.495\\
\texttt{P84126\_THETH} & 0.415 & 0.506 & 0.477 & 0.552 & 0.546 & 0.631 & 0.533 & 0.541 & 0.567\\
\texttt{PABP\_YEAST} & 0.640 & 0.666 & 0.667 & 0.693 & 0.665 & 0.662 & 0.641 & 0.689 & 0.639\\
\texttt{PA\_I34A1} & 0.456 & 0.493 & 0.533 & 0.538 & 0.054 & 0.383 & 0.541 & 0.572 & 0.539\\
\texttt{POLG\_CXB3N} & 0.328 & 0.382 & 0.374 & 0.369 & -0.057 & 0.476 & 0.347 & 0.405 & 0.465\\
\texttt{POLG\_HCVJF} & 0.390 & 0.434 & 0.443 & 0.487 & 0.605 & 0.600 & 0.525 & 0.577 & 0.614\\
\texttt{PTEN\_HUMAN} & 0.242 & 0.404 & 0.382 & 0.389 & 0.436 & 0.491 & 0.341 & 0.459 & 0.501\\
\texttt{Q2N0S5\_9HIV1} & 0.507 & 0.418 & 0.398 & 0.348 & 0.496 & 0.490 & 0.412 & 0.492 & 0.496\\
\texttt{Q59976\_STRSQ} & 0.579 & 0.638 & 0.644 & 0.654 & 0.506 & 0.674 & 0.645 & 0.659 & 0.647\\
\texttt{R1AB\_SARS2} & 0.214 & 0.259 & 0.274 & 0.289 & -0.030 & -0.037 & 0.216 & 0.401 & 0.600\\
\texttt{RASH\_HUMAN} & 0.439 & 0.419 & 0.423 & 0.399 & 0.359 & 0.415 & 0.377 & 0.447 & 0.454\\
\texttt{REV\_HV1H2} & 0.224 & 0.271 & 0.247 & 0.246 & 0.249 & 0.251 & 0.246 & 0.245 & 0.227\\
\texttt{RL401\_YEAST} & 0.384 & 0.522 & 0.488 & 0.436 & 0.288 & 0.392 & 0.355 & 0.397 & 0.395\\
\texttt{SC6A4\_HUMAN} & 0.420 & 0.410 & 0.400 & 0.411 & 0.472 & 0.509 & 0.400 & 0.465 & 0.489\\
\texttt{SCN5A\_HUMAN} & 0.121 & 0.151 & 0.167 & 0.131 & 0.176 & 0.157 & 0.073 & 0.093 & 0.199\\
\texttt{SPG1\_STRSG} & 0.252 & 0.216 & 0.214 & 0.208 & 0.237 & 0.142 & 0.279 & 0.289 & 0.247\\
\texttt{SPIKE\_SARS2} & 0.311 & 0.375 & 0.366 & 0.375 & -0.043 & 0.471 & 0.369 & 0.342 & 0.347\\
\texttt{SRC\_HUMAN} & 0.439 & 0.413 & 0.421 & 0.373 & 0.561 & 0.258 & 0.348 & 0.493 & 0.505\\
\texttt{SUMO1\_HUMAN} & 0.223 & 0.369 & 0.407 & 0.391 & 0.430 & 0.423 & 0.424 & 0.488 & 0.531\\
\texttt{SYUA\_HUMAN} & 0.195 & 0.219 & 0.195 & 0.186 & 0.281 & 0.128 & 0.160 & 0.146 & 0.167\\
\texttt{TADBP\_HUMAN} & 0.174 & 0.070 & -0.018 & -0.011 & 0.060 & 0.050 & 0.123 & 0.125 & 0.071\\
\texttt{TAT\_HV1BR} & 0.378 & 0.363 & 0.396 & 0.399 & 0.342 & 0.310 & 0.206 & 0.244 & 0.337\\
\texttt{TPK1\_HUMAN} & 0.099 & 0.151 & 0.261 & 0.289 & 0.284 & 0.268 & 0.313 & 0.314 & 0.230\\
\texttt{TPMT\_HUMAN} & 0.373 & 0.462 & 0.509 & 0.515 & 0.54 & 0.508 & 0.445 & 0.522 & 0.548\\
\texttt{TPOR\_HUMAN} & 0.245 & 0.327 & 0.305 & 0.369 & 0.362 & 0.507 & 0.410 & 0.453 & 0.393\\
\texttt{TRPC\_SACS2} & 0.425 & 0.558 & 0.499 & 0.568 & 0.606 & 0.629 & 0.551 & 0.585 & 0.577\\
\texttt{TRPC\_THEMA} & 0.333 & 0.392 & 0.392 & 0.452 & 0.472 & 0.474 & 0.453 & 0.436 & 0.420\\
\texttt{UBC9\_HUMAN} & 0.237 & 0.438 & 0.473 & 0.452 & 0.479 & 0.503 & 0.433 & 0.485 & 0.538\\
\texttt{UBE4B\_MOUSE} & 0.117 & 0.325 & 0.334 & 0.293 & 0.462 & 0.347 & 0.256 & 0.388 & 0.476\\
\texttt{VKOR1\_HUMAN} & 0.166 & 0.173 & 0.343 & 0.375 & 0.447 & 0.472 & 0.466 & 0.502 & 0.462\\
\texttt{YAP1\_HUMAN} & 0.180 & 0.177 & 0.160 & 0.168 & 0.281 & 0.071 & 0.218 & 0.359 & 0.438\\
\midrule
\texttt{Average} & 0.330 & 0.370 & 0.381 & 0.387 & 0.371 & 0.422 & 0.406 & 0.451 & 0.448\\
\bottomrule
\end{tabular}
\label{tab:full_fitness}
\end{tiny}
\end{center}
\end{table}
\begin{table}[H]
\caption{\textbf{Fitness evaluation - ProteinGym indel benchmark:} Spearman's rank correlation between experimentally measured fitness values for different proteins and the value predicted by the models. The Wavenet models are based on \citet{shinProteinDesignVariant2021}. Tranception NR and Tranception R \cite{Notin2022TranceptionPF} are variants without and with retrieval respectively.}
\begin{center}
\begin{tiny}
\begin{tabular}{lccccccc}
\toprule
Uniprot\_ID & \multicolumn{4}{c}{RITA}& \multicolumn{3}{c}{Baselines} \\
                                                 & \multicolumn{1}{c}{S} & \multicolumn{1}{c}{M} & \multicolumn{1}{c}{L} & \multicolumn{1}{c}{XL} & \multicolumn{1}{l}{Wavenet} &
                                                 \multicolumn{1}{l}{Tranception NR} &
                                                 \multicolumn{1}{l}{Tranception R} \\
\midrule
\texttt{A0A1J4YT16\_9PROT\_Davidi\_2020} & -0.169 & 0.185 & 0.207 & 0.210 & 0.117 & 0.178 & 0.191\\
\texttt{B1LPA6\_ECOSM\_Russ\_2020} & 0.292 & 0.383 & 0.339 & 0.348 & 0.385 & 0.321 & 0.415\\
\texttt{BLAT\_ECOLX\_Gonzalez\_indels\_2019} & 0.436 & 0.455 & 0.334 & 0.345 & 0.546 & 0.296 & 0.357\\
\texttt{CAPSD\_AAV2S\_Sinai\_indels\_2021} & 0.253 & 0.319 & 0.453 & 0.463 & 0.699 & 0.563 & 0.598\\
\texttt{HIS7\_YEAST\_Pokusaeva\_indels\_2019} & 0.638 & 0.656 & 0.677 & 0.684 & 0.457 & 0.549 & 0.586\\
\texttt{P53\_HUMAN\_Kotler\_deletions\_2018} & 0.360 & 0.407 & 0.383 & 0.273 & 0.680 & 0.707 & 0.692\\
\texttt{PTEN\_HUMAN\_Mighell\_deletions\_2018} & 0.612 & 0.575 & 0.504 & 0.523 & 0.001 & 0.395 & 0.401\\
\midrule
\texttt{Average} & 0.346 & 0.426 & 0.414 & 0.406 & 0.412 & 0.430 & 0.463\\
\bottomrule
\end{tabular}
\label{tab:full_fitness_indels}
\end{tiny}
\end{center}
\end{table}


\section{Positional embedding ablation}

\begin{table*}[h]
\caption{\textbf{Ablating different positional embeddings:} We evaluate Rotary Positional Embeddings (RoPE) as well as ALiBi by training a small model for 3 billion amino acids. As shown, RoPE outperform ALiBi. However, we note that the training runs were stopped after a short amount of time to save computational resources, and that larger scale ablation is needed for reliable results. We also note that the training dataset differs from the one used for the main training runs, and that for this reason these results should not be directly compared to those presented in Table \ref{tab:ppl}.}
\label{tab:rotary}
\vskip 0.15in
\begin{center}
\begin{small}
\begin{sc}
\begin{tabular}{l cc }
\toprule
 & Small-Rotary & Small-Alibi \\
\midrule
Perplexity & 12.43 & 13.08\\
\bottomrule
\end{tabular}
\end{sc}
\end{small}
\end{center}
\vskip -0.2in
\end{table*}

\section{Dataset Selection}

\begin{table*}[h]
\caption{\textbf{Ablating different datasets:} We train small models for $\sim3$ GT to evaluate the use of different pre-training datasets. All model are then evaluated on a combination of the datasets, and the results are shown below.}
\label{tab:dataset_selection}
\vskip 0.15in
\begin{center}
\begin{small}
\begin{sc}
\begin{tabular}{l ccc }
\toprule
 & Uniref-100 & MetaClust & MGnify \\
\midrule
Perplexity & 14.28	& 14.62	& 15.34\\
\bottomrule
\end{tabular}
\end{sc}
\end{small}
\end{center}
\vskip -0.1in
\end{table*}
\section{Comparisons with ProtXLNet}

A previous version of this paper contained faulty perplexity comparisons with ProtXLNet, after correspondence with the authors we have decided to entirely remove comparisons with ProtXLNet from the main part of the paper, and provide a fixed perplexity evaluation here, in the appendix . The goal of these comparisons was to contextualize our models compared to similar previous models. However, the XLNet architecture is rather different to that of a decoder-only autoregressive transformer. In table~\ref{tab:ppl-xlnet} we show the perplexity evaluation, with correctly computed values for ProtXLNet. Additionally in this evaluation we have also removed all proteins of length less than one hundred amino acids.

\begin{table*}[h]
\caption{\textbf{Perplexity evaluation:} We evaluate generative protein models on the upstream modeling task by measuring the models perplexity-per-byte on four different datasets. In all cases performance is correlated with model size and RITA-XL provides the best results, highlighted in \textbf{bold}. We further note that both ProtGPT2 and ProtXLNet were trained on the Pfam families we held out of our training set.}
\label{tab:ppl-xlnet}
\vskip 0.15in
\begin{center}
\begin{small}
\begin{sc}
\begin{tabular}{lcccc cc}
\toprule
& \multicolumn{4}{c}{RITA}& \multicolumn{2}{c}{Baselines}\\

Dataset & Small & Medium & Large & XLarge & ProtGPT2 & ProtXLNet \\
\midrule
UniRef-100    & 9.93	& 7.32	& 6.05	& \textbf{5.36}	& 17.96	& 18.32 \\
Metaclust     & 12.93	& 11.26	& 9.93	& \textbf{9.33}	& 20.16	& 22.33 \\
MGnify        & 12.90	& 11.33&  10.03	& \textbf{9.21}	& 19.97	& 22.52 \\
Pfam heldout  & 11.76	& 10.66	& 9.21	& \textbf{7.93}	& 15.02	& 16.39 \\
\bottomrule
\end{tabular}
\end{sc}
\end{small}
\end{center}
\vskip -0.1in
\end{table*}

\section{Comparing perplexity across different vocabularies}
\label{sec:ppl_per_byte}
Evaluating the perplexity of a model has long been standard practice in natural language processing. However, the traditional way of computing the perplexity, per token, does not transfer across vocabularies. It is naturally much harder to predict the correct next token in a large vocabulary of tens of thousands of tokens compared to a small vocabulary with only a few dozens tokens to chose from.

In order to compare across vocabularies one must normalize the perplexity by the length of the untokenized sequence, instead of the tokenized sequence. This metric is typically called the \textit{perplexity per byte}, although for proteins it may be more natural to call it the perplexity per amino-acid. If one merges tokens and assigns the probability of the merged tokens as the joint probability of its constituents, the perplexity per byte will remain constant. This property allows fair comparisons across vocabulary sizes. However, perplexity per byte also has its flaws: in redundant vocabularies, where each sequence can be represented in multiple ways, the perplexity per byte will unjustly increase. In such a case the model needs to guess which of all possible tokenizations corresponds to the target sequence. See Table~\ref{tab:ppl_vocabs} for an example.

While all common tokenizers are deterministic, meaning that the model should be able to learn which of the possible tokenizations the tokenizer will chose, we hypothesize that this can be a difficult task when the length of each word grows longer. For proteins, where there are no word boundaries, this may cause problems when using standard tokenizers, such as BPE. In preliminary experiments using tokenization we saw improved results by breaking up the proteins into k-mers before tokenization.

While perplexity per byte is the standard way to compare upstream performance across vocabularies we would like to caution against reading too much into the exact values. Unfortunately, we are not aware of any better metric to compare across vocabularies. 

\begin{table*}[h]
\caption{\textbf{Perplexity and vocabularies}: Example of how the perplexity per token and the perplexity per byte change with vocabularies given a uniform probabilities across the tokens in the vocabulary.}
\label{tab:ppl_vocabs}
\vskip 0.15in
\begin{center}
\begin{small}
\begin{sc}
\begin{tabular}{c c c c c }
\toprule
Vocab & Vocab type & Sequence & Perplexity per token & Perplexity per byte\\
\midrule
A, B &  Untokenized  & A,B,B,A& $\exp(-\frac{4*ln(0.5)}{4})=2$& $\exp(-\frac{4*ln(0.5)}{4})=2$ \\
AA, BB, AB, BA & Tokenized &AB, BA & $\exp(-\frac{2*ln(0.25)}{2})=4$& $\exp(-\frac{2*ln(0.25)}{4})=2$\\
A, B, AA, BB, AB, BA & Redundantly tokenized &A, BB, A & $\exp(-\frac{3*ln(0.167)}{3})=6$& $\exp(-\frac{3*ln(0.167)}{4})\approx3.83$\\
\bottomrule
\end{tabular}
\end{sc}
\end{small}
\end{center}
\vskip -0.1in
\end{table*}

\section{Acknowledgments}

The authors thank Julien Launay and Igor Carron for fruitful discussions.

This work was made possible through the use of the HPC/AI resources at IDRIS under GENCI allocation 2020-AD011012024 which provided access to the Jean Zay supercomputer. We thank Stéphane Réquena and the support team for their valuable help.\\

Daniel Hesslow is supported by the European Union’s Horizon 2020 research and innovation programme under the Marie Sk\l{}odowska-Curie grant agreement No 860360.\\
Pascal Notin is supported by GSK and the UK Engineering and Physical Sciences Research Council (EPSRC ICASE award no. 18000077). \\
\begin{center}
    \includegraphics[width = 0.12\textwidth]{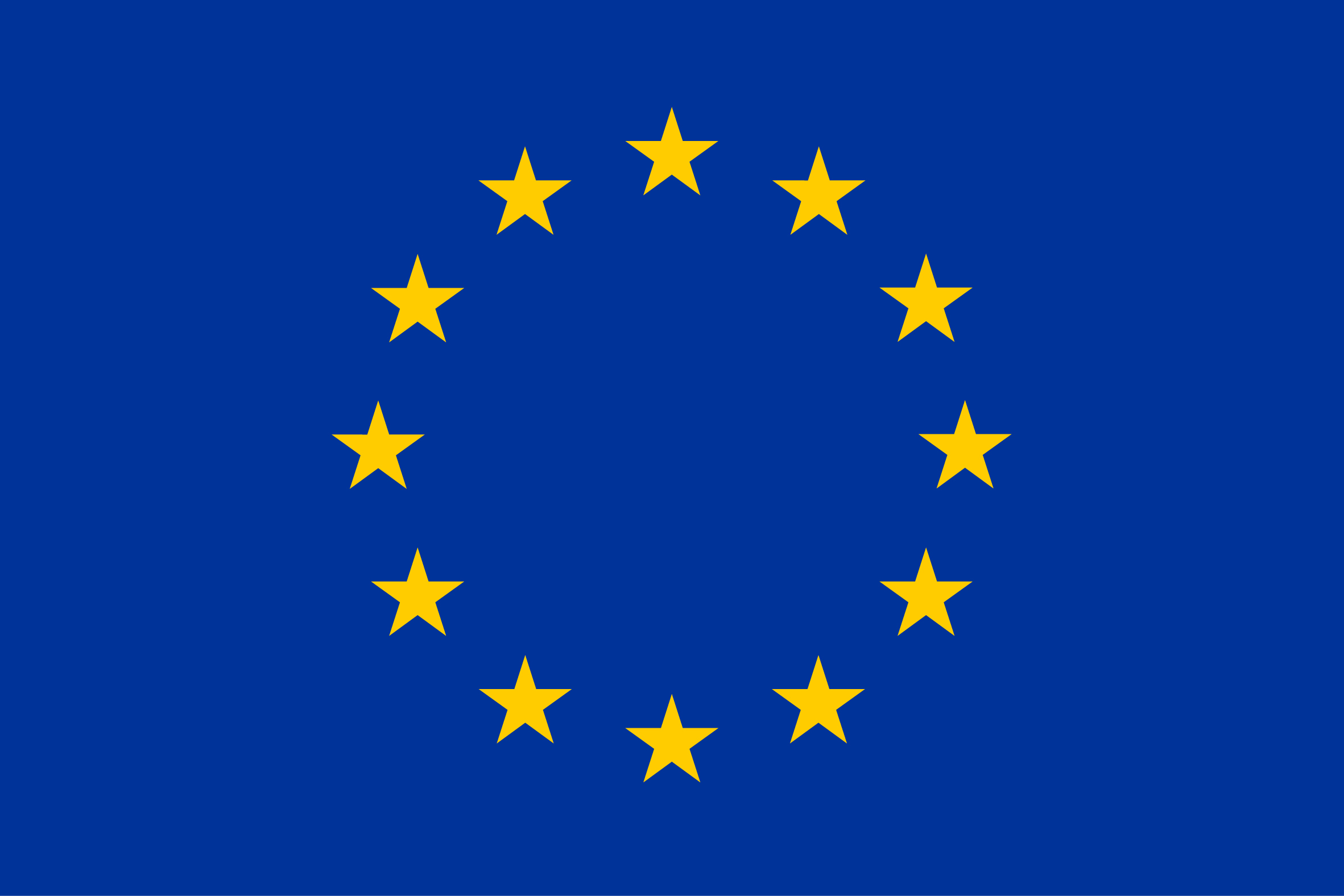}\\
\end{center}

\end{document}